\documentclass[prd,twocolumn,showpacs,preprintnumbers,amsmath,amssymb,superscriptaddress,floatfix,nofootinbib]{revtex4}

\usepackage{graphicx}
\usepackage{amsmath}
\usepackage{amsfonts}
\usepackage{amssymb}
\usepackage{color}
\usepackage{multirow}

\newcommand{\Slash}[1]{\ooalign{\hfil/\hfil\crcr$#1$}}

\begin{document}

\title{Neutral hidden charm pentaquark states $P_c^0(4380)$ and $P_c^0(4450)$ in $\pi^-p \to J/\psi n$ reaction}

\author{Qi-Fang L\"{u}} \email{lvqifang@ihep.ac.cn}
\affiliation{Institute of High Energy Physics, Chinese Academy of Sciences, Beijing 100049, China}
\author{Xiao-Yun Wang}
\email{xywang@impcas.ac.cn}
\affiliation{Institute of Modern Physics, Chinese Academy of Sciences, Lanzhou 730000, China}
\affiliation{University of Chinese Academy of Sciences, Beijing 100049, China}
\affiliation{Research Center for Hadron and CSR Physics, Institute of Modern Physics of
CAS and Lanzhou University, Lanzhou 730000, China}
\author{Ju-Jun Xie} \email{xiejujun@impcas.ac.cn}
\affiliation{Institute of Modern Physics, Chinese Academy of Sciences, Lanzhou 730000, China}
\affiliation{Research Center for Hadron and CSR Physics, Institute of Modern Physics of
CAS and Lanzhou University, Lanzhou 730000, China}
\affiliation{State Key Laboratory of Theoretical Physics, Institute of Theoretical
Physics, Chinese Academy of Sciences, Beijing 100190, China}
\author{Xu-Rong Chen} \email{xchen@impcas.ac.cn}
\affiliation{Institute of Modern Physics, Chinese Academy of Sciences, Lanzhou 730000, China}
\affiliation{Research Center for Hadron and CSR Physics, Institute of Modern Physics of
CAS and Lanzhou University, Lanzhou 730000, China}
\author{Yu-Bing Dong} \email{dongyb@ihep.ac.cn}
\affiliation{Institute of High Energy Physics, Chinese Academy of Sciences, Beijing 100049, China}
\affiliation{Theoretical Physics Center for Science Facilities (TPCSF), CAS, Beijing 100049, China}

\begin{abstract}

We investigate the neutral hidden charm pentaquark states $P_c^0(4380)$ and $P_c^0(4450)$ in $\pi^-p \to J/\psi n$ reaction within an effective Lagrangian approach. The background contributions for the process mainly come from $t$-channel $\pi$ and $\rho$ meson exchanges. The contributions of $P_c^0(4380)$ and $P_c^0(4450)$ states give clear peak structures in the magnitude of 1 $\mu$b at center of mass energy 4.38 GeV and 4.45 GeV in the total cross sections. Hence, this reaction may provide a new good platform to search for neutral $P_c$ states. It is expected that our estimated total cross sections together with the angular distributions can be tested by future experiments at J-PARC.
\end{abstract}
\maketitle

\section{Introduction}{\label{introduction}}

Recently, LHCb Collaboration observed two exotic structures in the $J/\psi p$ invariant mass spectrum in the $\Lambda_b^0 \to J/\psi K^- p$ process\cite{Aaij:2015tga}. The lower state $P_c^+(4380)$, has a mass of $4380\pm8\pm29$ MeV and a width of $205\pm18\pm86$ MeV, while the mass and width of the higher state $P_c^+(4450)$ are $4449.8\pm1.7\pm2.5$ MeV and $39\pm5\pm19$ MeV, respectively. Three pairs of possible spin-parity values are favored for $P_c^+(4380)$ and $P_c^+(4450)$, which are $(3/2^-, 5/2^+)$, $(3/2^+, 5/2^-)$, and $(5/2^+, 3/2^-)$.

Very recently, LHCb Collaboration reported the branching fraction of the decay $\Lambda_b^0 \to J/\psi K^- p$\cite{Aaij:2015fea}. Together with the fractions of $P_c^+(4380)$ and $P_c^+(4450)$ in $\Lambda_b^0 \to J/\psi K^- p$ decay measured previously, the branching ratios ${\cal B}(\Lambda_b^0 \to P_c^+ K^-){\cal B}(P_c^+ \to J/\psi p)$ are determined as

\begin{eqnarray}
&&{\cal B}(\Lambda_b^0 \to P_c^+(4380) K^-){\cal B}(P_c^+(4380) \to J/\psi p) \nonumber \\
&&= 2.56\pm0.22\pm1.28^{+0.46}_{-0.36} \times 10^{-5},
\end{eqnarray}

\begin{eqnarray}
&&{\cal B}(\Lambda_b^0 \to P_c^+(4450) K^-){\cal B}(P_c^+(4450) \to J/\psi p) \nonumber \\
&&= 1.25\pm0.15\pm0.33^{+0.22}_{-0.18} \times 10^{-5}.
\end{eqnarray}

The observations immediately attract lots of theoretical works on these two states. Various interpretations, such as loosely bound molecular states\cite{Karliner:2015ina,Chen:2015loa,Roca:2015dva,He:2015cea,Chen:2015moa,Meissner:2015mza}, compact pentaquark states\cite{Maiani:2015vwa,Wang:2015epa,Anisovich:2015xja,Wang:2015ava,Wang:2015wsa,Lebed:2015tna,Li:2015gta}, and anomalous triangle singularity effects\cite{Guo:2015umn,Liu:2015fea,Mikhasenko:2015vca}, are respectively proposed. Those calculations mainly focus on the masses of $P_c$ states, and a comprehensive discussion of various interpretations can be found in Ref.~\cite{Burns:2015dwa}. It should be noted that the hidden charm states have already been investigated by many works with meson-baryon and meson-meson interactions in the literatures, in which the masses and decay widths are calculated\cite{Wu:2010jy,Wu:2010vk,Wu:2010rv,Wu:2012md, Xiao:2013yca,Uchino:2015uha,Zou:2013af,Yang:2011wz,Wang:2011rga}. These states are probably the partners of the observed $P_c$ states.

Besides the static properties, the production mechanism of $P_c$ states is also an important topic. There have been some studies on the production of hidden charm states before the observations of $P_c$ states, in which only lower spin states are considered\cite{Wu:2012xg,Wu:2012wta,Huang:2013mua,Garzon:2015zva,Wang:2015qia,Wang:2015xwa}. In $\Lambda_b^0 \to J/\psi K^- p$ decay process, only charged $P_c$ states can be observed. The $\gamma p \to J/\psi p$ reaction with charged $P_c$ production is proposed by some theoretical works and is expected to be tested by JLab experiment in the near future\cite{Wang:2015jsa,Kubarovsky:2015aaa,Karliner:2015voa}. However, there is few study on the production of its neutral partners in the literature. In Refs.~\cite{Lebed:2015tna,Cheng:2015cca}, the authors suggest that the neutral $P_c$ states can be produced via $\Lambda_b \to J/\psi \bar{K}^0 n$ decay process. This situation is different from the studies of $Z_c$ family, where both the charged and neutral $Z_c(3900)$, $Z_c(4020)$, $Z_c(4200)$, and $Z_c(4430)$ etc., have been discussed and analyzed in detail both experimentally and theoretically\cite{Liu:2008qx,Lin:2013mka,Ablikim:2013mio,Aaij:2014jqa,Xiao:2013iha,Ablikim:2014dxl,Wang:2015uua,Wang:2015pfa}. It is of great interest to search for the neutral $P_c$ states in addition to the charged ones. We expect that the analyses of the
$\pi^-p \to J/\psi n$ reaction at J-PARC could give information about the neutral ones and therefore, provide a unique perspective to the nature of hidden charm $P_c$ states.

In the present work, we study the production of neutral $P_c$ states in the pion induced reaction with an effective Lagrangian approach. There have been several papers related to the exotic resonances at J-PARC\cite{Garzon:2015zva,Wang:2015qia,Wang:2015xwa,Xie:2015zga}. In this pion beam experiments at J-PARC, the expected pion energy can reach up to 20 GeV in the laboratory frame\cite{Kim:2014qha} with high luminosity, which is enough to product the $P_c$ states via $\pi^- p$ collision, and therefore, the
measurement at J-PARC can test our calculations particular for the neutral $P_c$ states.

This paper is organized as follows. In Sec.\ref{formalisms}, the formalisms and ingredients for our calculations are listed. The results of total and differential cross sections and discussions are presented in Sec.\ref{results}. Finally, a short summary is given in the last section.

\section{Formalisms and Ingredients}{\label{formalisms}}

\begin{figure*}[htbp]
\includegraphics[scale=0.5]{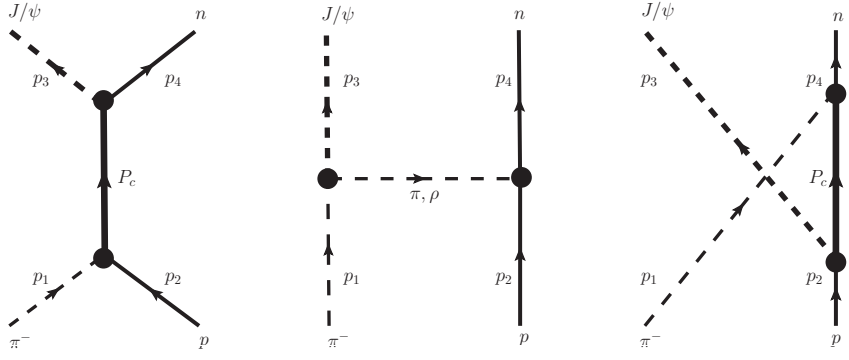}
\vspace{0.0cm} \caption{Feynman diagrams for $\pi^- p \to J/\psi n$ reaction.}
\label{fey}
\end{figure*}

Here we study the $\pi^- p \to J/\psi n$ reaction within an effective Lagrangian approach, which has been widely employed to investigate the pion induced reactions\cite{Garzon:2015zva,Kim:2014qha,Xie:2007qt,Lu:2013jva,Xie:2015zga,Wu:2014yca,Xie:2014kja,Wang:2014ofa,Kim:2015ita}. The relevant Feynman diagrams are depicted in Fig.~\ref{fey}. The $s$-channel $P_c$ states with different spin-parity assumptions are involved in our analyses. The $u$-channel contributions are expected to be negligible due to the highly off-shell intermediate $P_c$ states. The background contributions from $t$-channel via $\pi$ and $\rho$ meson exchanges are taken into account, while other
meson exchanges in the $t$-channel, such as $Z_c^0(3900)$, are simply ignored due to their unclear structures.

The effective Lagrangians for $P_cNJ/\psi$ couplings can be written as\cite{Kim:2011rm}:

\begin{eqnarray}
{\cal L}_{P_cNJ/\psi}^{{3/2}^\pm}&=&-\frac{i g_1}{2M_N}\bar N \Gamma_{\nu}^{(\pm)}\psi^{\mu\nu}P_{c\mu}\nonumber \\
&&-\frac{g_2}{(2M_N)^2}\partial_{\nu}\bar N \Gamma^{(\pm)}\psi^{\mu\nu}P_{c\mu}\nonumber \\
&&+\frac{g_3}{(2M_N)^2}\bar N \Gamma^{(\pm)}\partial_{\nu}\psi^{\mu\nu}P_{c\mu} + \rm{H.c}.,
\end{eqnarray}

\begin{eqnarray}
{\cal L}_{P_cNJ/\psi}^{{5/2}^\pm}&=&\frac{g_1}{(2M_N)^2}\bar N \Gamma_{\nu}^{(\mp)}\partial^{\alpha}\psi^{\mu\nu}P_{c\mu\alpha}\nonumber \\
&&-\frac{i g_2}{(2M_N)^3}\partial_{\nu}\bar N \Gamma^{(\mp)}\partial^{\alpha}\psi^{\mu\nu}P_{c\mu\alpha}\nonumber \\
&&+\frac{i g_3}{(2M_N)^3}\bar N \Gamma^{(\mp)}\partial^{\alpha}\partial_{\nu}\psi^{\mu\nu}P_{c\mu\alpha} + \rm{H.c}.,
\end{eqnarray}
where the vertex $\Gamma$ matrix is defined as:

\begin{equation}
\Gamma_{\mu}^{(\pm)} \equiv \left(
 \begin{array}{cr}
\gamma_{\mu}\gamma_5 \\
\gamma_{\mu}
\end{array}
\right),
\end{equation}

\begin{equation}
\Gamma^{(\pm)} \equiv \left(
 \begin{array}{cr}
\gamma_5 \\
1
\end{array}
\right),
\end{equation}
for positive and negative parities.

In our calculation, the higher partial wave terms in eqs. (3-4)  are neglected due to the small momentum of the final $J/\psi N$ state compared with nucleon mass and due to the lack of experimental data\cite{Wang:2015jsa}. Therefore, we have only one unknown coupling $g_1$. The effective Lagrangians for $P_cN \pi$ vertexes are described in a Lorentz covariant orbital-spin scheme\cite{Zou:2002yy}:

\begin{eqnarray}
{\cal L}_{P_cN\pi}^{{3/2}^+}&=&\frac{g_{P_cN\pi}}{m_{\pi}}\bar N \vec \tau \cdot \partial_{\mu} \vec \pi P_c^{\mu}+ \rm{H.c}.,
\end{eqnarray}

\begin{eqnarray}
{\cal L}_{P_cN\pi}^{{3/2}^-}&=&\frac{g_{P_cN\pi}}{m_{\pi}^2}\bar N \gamma_5 \gamma_{\mu} \vec \tau \cdot \partial^{\mu} \partial_{\nu} \vec \pi P_c^{\nu}+ \rm{H.c}.,
\end{eqnarray}

\begin{eqnarray}
{\cal L}_{P_cN\pi}^{{5/2}^+}&=&\frac{g_{P_cN\pi}}{m_{\pi}^3}\bar N \gamma_5 \gamma_{\mu} \vec \tau \cdot \partial^{\mu} \partial_{\nu} \partial_{\lambda} \vec \pi P_c^{\nu\lambda}+ \rm{H.c}.,
\end{eqnarray}

\begin{eqnarray}
{\cal L}_{P_cN\pi}^{{5/2}^-}&=&\frac{g_{P_cN\pi}}{m_{\pi}^2}\bar N \vec \tau \cdot \partial_{\mu} \partial_{\nu} \vec \pi P_c^{\mu\nu}+ \rm{H.c}..
\end{eqnarray}

Furthermore, we adopt the commonly used Lagrangian densities for $J/\psi \pi \pi$, $J/\psi \pi \rho$, $\pi N N$, and $\rho N N$ vertexes\cite{Liang:2004sd,Lu:2014rla,Ryu:2012tw,Wu:2013xma,Xie:2007qt} in our $t$-channel calculation. They are

\begin{eqnarray}
{\cal L}_{J/\psi \pi \pi}&=& -i g_{J/\psi \pi \pi}(\partial^{\mu} \pi^- \pi^+ - \partial^{\mu} \pi^+ \pi^-)\psi_{\mu},
\end{eqnarray}

\begin{eqnarray}
{\cal L}_{J/\psi \pi \rho}&=& - \frac{g_{J/\psi\pi\rho}}{m_{J/\psi}} \varepsilon^{\mu\nu\alpha\beta} \partial_{\mu} \rho_\nu \partial_\alpha \psi_\beta \pi,
\end{eqnarray}

\begin{eqnarray}
{\cal L}_{\pi N N}&=&-\frac{g_{\pi NN}}{2M_N}\bar N \gamma_5 \gamma_{\mu} \vec \tau \cdot \partial^{\mu} \vec \pi N,
\end{eqnarray}

\begin{eqnarray}
{\cal L}_{\rho N N}&=&-g_{\rho NN} \bar N (\gamma_{\mu} + \frac{\kappa}{2M_N} \sigma_{\mu\nu} \partial^\nu) \vec \tau \cdot \partial^{\mu} \vec \rho N.
\end{eqnarray}

In this work, the decay processes of $P_c \to NJ/\psi$ and $P_c \to N\pi$ are calculated and the relevant coupling constants $g_1(\equiv g_{P_cNJ/\psi})$ and $g_{P_c N\pi}$ can be obtained from their partial decay widths with different $J^P$ assignments of $P_c$. The obtained coupling constants are listed in Table  \ref{tab}, by assuming that the branching ratios are $10\%$ and $1\%$ for the $P_c \to NJ/\psi$ and $P_c \to N\pi$, respectively. In the calculation,  we employ the total widths of the two $P_c$ states from experimental
measurements with  $\Gamma_{Pc(4380)}= 205~\rm{MeV}$ and $\Gamma_{P_c(4450)}= 39~\rm{MeV}$ .

\begin{table}
\begin{center}
\caption{ \label{tab} Coupling constants of $P_cNJ/\psi$ and $P_cN\pi$ different $J^P$ assignments by assuming the branching ratios are $10\%$ and $1\%$, respectively.}
\scriptsize
\begin{tabular}{lccccc}
\hline\hline
  State          & Channel            & $3/2^+$       & $3/2^-$   & $5/2^+$     &$5/2^-$ \\\hline
  $P_c(4380)$    & $J/\psi N$         & 1.09          &0.49       & 2.17        &5.13  \\
                 & $\pi N$            & $8.56\times 10^{-3}$      &  $3.43\times 10^{-4}$     & $3.59\times 10^{-5}$        &$8.95\times 10^{-4}$  \\
  $P_c(4450)$    & $J/\psi N$         & 0.41          &0.20       & 0.80        &1.75  \\
                 & $\pi N$            & $3.65\times 10^{-3}$      &  $1.43\times 10^{-4}$     & $1.47\times 10^{-5}$        &$3.75\times 10^{-4}$  \\
\hline\hline
\end{tabular}
\end{center}
\end{table}

The coupling constants of $J/\psi \pi \rho$, $J/\psi \pi \pi$, $\pi NN$ and $\rho NN$ are needed as well in our calculation, and we select $g_{J/\psi \pi \pi} = 8.20 \times 10^{-4}$, $g_{J/\psi \pi \rho} = 0.032$, $g_{\pi NN}=13.45$, $g_{\rho NN}^2/(4\pi) = 0.9$ and $\kappa=6.1$ according to Refs. \cite{Wu:2013xma,Xie:2007qt}.

The propagators for exchanged $\pi$ and $\rho$ mesons are

\begin{eqnarray}
G_{\pi}(q)&=& \frac{i}{q^2-m^2_\pi},
\end{eqnarray}

\begin{eqnarray}
G_{\rho}^{\mu\nu}(q)&=& i\frac{-g^{\mu\nu}+q^\mu q^\nu / m_{\rho}^2}{q^2-m^2_\rho}.
\end{eqnarray}

For the propagator of spin-3/2 fermion, we use

\begin{eqnarray}
G^{\beta\alpha}(q)&=& \frac{i (\Slash{q}+M) P^{\beta\alpha}(q)}{q^2-M^2+i M \Gamma},
\end{eqnarray}
with

\begin{eqnarray}
P^{\beta\alpha}(q)&=& -g^{\beta\alpha}+\frac{1}{3}\gamma^\beta \gamma^\alpha + \frac{1}{3M} (\gamma^\beta q^\alpha - \gamma^\alpha q^\beta) \nonumber \\
&&+ \frac{2}{3M^2}q^\beta q^\alpha,
\end{eqnarray}
and for the propagator of spin-5/2 fermion, it is

\begin{eqnarray}
G^{\rho\sigma\alpha\beta}(q)&=& \frac{i (\Slash{q}+M) P^{\rho\sigma\alpha\beta}(q)}{q^2-M^2+i M \Gamma},
\end{eqnarray}
with

\begin{eqnarray}
P^{\rho\sigma\alpha\beta}(q)&=& \frac{1}{2}(\tilde g^{\rho\alpha} \tilde g^{\sigma\beta} + \tilde g^{\rho\beta} \tilde g^{\sigma\alpha}) -\frac{1}{5} \tilde g^{\rho\sigma}\tilde g^{\alpha\beta} \nonumber \\
&& -\frac{1}{10} (\tilde \gamma^\rho \tilde \gamma^\alpha \tilde g^{\sigma\beta} + \tilde \gamma^\rho \tilde \gamma^\beta \tilde g^{\sigma\alpha} \nonumber \\
&&+ \tilde \gamma^\sigma \tilde \gamma^\alpha \tilde g^{\rho\beta}+\tilde \gamma^\sigma \tilde \gamma^\beta \tilde g^{\rho\alpha}),
\end{eqnarray}
where
\begin{eqnarray}
\tilde g^{\alpha\beta}&=& g^{\alpha\beta} - \frac{p^\alpha p^\beta}{M^2},
\end{eqnarray}
and
\begin{eqnarray}
\tilde \gamma^{\alpha}&=& \gamma^{\alpha} - \frac{p^\alpha}{M^2}\Slash{p}.
\end{eqnarray}

From the above Lagrangian densities, the $s$-channel amplitude for each $J^P$ assignment of $P_c$ states   can be obtained,

\begin{eqnarray}
{\cal M}^{3/2^+}&=& \frac{i g_{P_cNJ/\psi}}{2M_N} \frac{\sqrt{2} g_{P_cN\pi}}{m_{\pi}} F(q^2) \epsilon^*_{\nu}(p_3,s_3) \nonumber \\
&& \bar u (p_4,s_4) \gamma_\sigma \gamma_5 (p_3^\beta g^{\nu\sigma}-p_3^\sigma g^{\beta\nu})\nonumber \\
&& G_{\beta \alpha}(q) p_1^{\alpha} u(p_2,s_2),
\end{eqnarray}

\begin{eqnarray}
{\cal M}^{3/2^-}&=& \frac{i g_{P_cNJ/\psi}}{2M_N} \frac{-\sqrt{2}i g_{P_cN\pi}}{m_{\pi}^2} F(q^2) \epsilon^*_{\nu}(p_3,s_3) \nonumber \\
&& \bar u (p_4,s_4) \gamma_\sigma (p_3^\beta g^{\nu\sigma}-p_3^\sigma g^{\beta\nu})\nonumber \\
&& G_{\beta \alpha}(q)\gamma_5\Slash {p_1} p_1^{\alpha} u(p_2,s_2),
\end{eqnarray}

\begin{eqnarray}
{\cal M}^{5/2^+}&=& \frac{-i g_{P_cNJ/\psi}}{(2M_N)^2} \frac{-\sqrt{2} g_{P_cN\pi}}{m_{\pi}^3} F(q^2) \epsilon^*_{\nu}(p_3,s_3)\nonumber \\
&& \bar u (p_4,s_4)\gamma_\delta p_3^\sigma (p_3^\rho g^{\nu\delta}-p_3^\delta g^{\rho\nu})\nonumber \\
&&G_{\rho\sigma\alpha\beta}(q)\gamma_5\Slash {p_1} p_1^{\alpha} p_1^{\beta} u(p_2,s_2),
\end{eqnarray}

\begin{eqnarray}
{\cal M}^{5/2^-}&=& \frac{-i g_{P_cNJ/\psi}}{(2M_N)^2} \frac{-\sqrt{2}i g_{P_cN\pi}}{m_{\pi}^2} F(q^2) \epsilon^*_{\nu}(p_3,s_3)\nonumber \\
&& \bar u (p_4,s_4)\gamma_\delta \gamma_5 p_3^\sigma (p_3^\rho g^{\nu\delta}-p_3^\delta g^{\rho\nu})\nonumber \\
&&G_{\rho\sigma\alpha\beta}(q) p_1^{\alpha} p_1^{\beta} u(p_2,s_2).
\end{eqnarray}
Here $p_1$, $p_2$, $p_3$, and $p_4$ are the four momenta of pion, proton, $J/\psi$, and neutron, respectively; $s_2$, $s_3$, and $s_4$ are the spin projections of proton, $J/\psi$, and neutron, respectively. $q=p_1+p_2$ is the four momentum of the intermediate $P_c$ states.

In addition, the background $t$-channel $\pi$ and $\rho$ meson exchange amplitudes are

\begin{eqnarray}
{\cal M}_{\pi}&=& \frac{\sqrt{2} i g_{J/\psi \pi \pi} g_{\pi NN}}{M_N} F_{\pi}^{NN}(q_{\pi}^2) F_{\pi}^{J/\psi \pi}(q_{\pi}^2) \epsilon^*_{\nu}(p_3,s_3) \nonumber \\
&& p_1^{\nu} G_{\pi}(q) \bar u (p_4,s_4) \gamma_5 \Slash{q_{\pi}} u(p_2,s_2).
\end{eqnarray}

\begin{eqnarray}
{\cal M}_{\rho}&=& \frac{\sqrt{2} g_{J/\psi \pi \rho} g_{\rho NN}}{M_{J/\psi}} F_{\rho}^{NN}(q_{\rho}^2) F_{\rho}^{J/\psi \pi}(q_{\rho}^2) \epsilon^*_{\nu}(p_3,s_3)\nonumber \\
&& \varepsilon^{\alpha\beta\mu\nu} q_{\rho\alpha} p_{3\mu} G_{\rho\beta\lambda}(q)\bar u (p_4,s_4) \nonumber \\
&&[\gamma^\lambda+\frac{\kappa}{4M_N}(\gamma^\lambda \Slash{q_\rho} -\Slash{q_\rho} \gamma^\lambda)] u(p_2,s_2),
\end{eqnarray}
where $q_{\pi}=p_1-p_3$ and $q_{\rho}=p_1-p_3$ are the four momentum of $\pi$ and $\rho$ mesons, respectively.

In our calculations, phenomenological form factors are need since the hadrons are not point-like particles. Those form factors $F(q^2)$, $F_M^{NN}(q_M^2)$, and $F_M^{J/\psi \pi}(q_M^2)$ can be expressed as

\begin{eqnarray}
F(q^2)&=& \frac{\Lambda_{P_c}^4}{\Lambda_{P_c}^4 + (q^2-M_{P_c}^2)^2},
\end{eqnarray}

\begin{eqnarray}
F_M^{J/\psi \pi}(q_M^2) &=& \frac{\Lambda_M^{*2} -m_M^2}{\Lambda_M^{*2} -q_M^2}.
\end{eqnarray}

\begin{eqnarray}
F_M^{NN}(q_M^2) &=& (\frac{\Lambda_M^2 -m_M^2}{\Lambda_M^2 -q_M^2})^n.
\end{eqnarray}
with $n=1$ for $\pi$ meson and $n=2$ for $\rho$ meson\cite {Xie:2007qt}. We use the cutoff parameters $\Lambda_{P_c} = 0.5$ GeV for $P_c$ states\cite{Wang:2015jsa,Kim:2011rm}, and $\Lambda_{\rho}^* = \Lambda_{\pi}^* =1.3 $ GeV, $\Lambda_{\rho} = 1.6 $ GeV, $\Lambda_{\pi} =1.3 $ GeV for mesons\cite{Xie:2007qt}.

The unpolarized differential cross section in the c.m. frame for $\pi^-p \to J/\psi n$ reaction is

\begin{eqnarray}
\frac{d\sigma}{dcos\theta} &=& \frac{M_N^2}{16\pi s} \frac{|{\vec p}_3^{~c.m.}|}{|{\vec p}_1^{~c.m.}|}|{\cal M}_{\pi^-p \to J/\psi n}|^2.
\end{eqnarray}
with $\theta$ is the scattering angle of outgoing $J/\psi$ relative to the incoming pion beam, and ${\vec p}_1^{~c.m.}$ and ${\vec p}_3^{~c.m.}$ are the three momenta of $\pi$ and $J/\psi$ mesons in c.m.~frame. The relative phases between different amplitudes are unknown\cite{Lu:2014yba}. The interference terms with different choices of relative phases are calculated and these theoretical uncertainties for the total cross sections are presented.

\section{Results and Discussions}{\label{results}}

\begin{figure*}[htbp]
\includegraphics[scale=1.5]{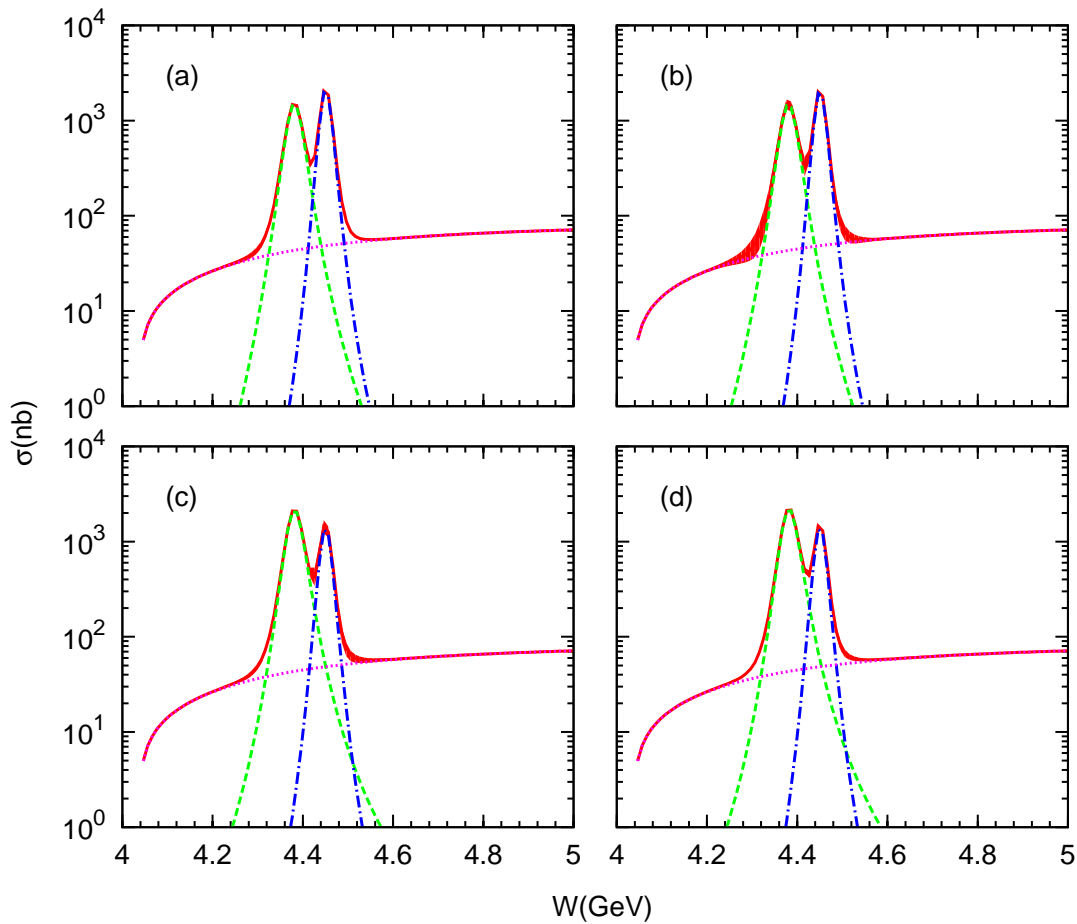}
\vspace{0.0cm} \caption{(Color online) The total cross sections for $\pi^- p \to J/\psi n$ reaction with different $J^P$ assumptions versus c.m. energy. The green dashed, blue dot-dashed, and pink short dotted lines stand for $P_c^0(4380)$, $P_c^0(4450)$, and background contributions, respectively. The thin red solid bands are total cross sections with the consideration of the interferences. (a), (b), (c), and (d) correspond to $(3/2^+,5/2^-)$, $(3/2^-,5/2^+)$, $(5/2^+,3/2^-)$, $(5/2^-,3/2^+)$ assumptions for $(P_c^0(4380),P_c^0(4450))$, respectively.}
\label{total}
\end{figure*}

Fig.~\ref{total} gives the total cross sections for $\pi^- p \to J/\psi n$ reaction with different $J^P$ assignments from threshold up to 5 GeV of the c.m.~energy. Besides the $t$-channel $\pi$ and $\rho$ meson exchanges, the $s$-channel $P_c^0(4380)$ and $P_c^0(4450)$ contributions are explicitly  presented. In the figure, the green dashed, blue dot-dashed, and pink short dotted lines stand for $P_c^0(4380)$, $P_c^0(4450)$, and background contributions, respectively. The red solid bands stand for the total cross sections due to the unknown relative phases between different amplitudes. It should be noted that the $(5/2^-,3/2^+)$ assumption for $(P_c(4380),P_c(4450))$ shown in Fig.~\ref{total} (d) is not favored by experiments\cite{Aaij:2015tga}. The $P_c$ states are firstly observed in the $J/\psi p$ invariant mass, and the $P_c \to J/\psi N$ decay processes can occur via falling apart mechanism. For $\pi N$ decay channel, these processes are OZI-allowed and two-body strong decays with large phase spaces. Large decay branching ratios are expected, if no $c\bar{c}$ pair annihilation is considered. The suppression due to the $c\bar{c}$ pair annihilation can be estimated by a factor of $(m_u/m_c)^2$, where $m_u$ and $m_c$ are the constituent quark masses of light quark and charm quark, respectively. This assumption has been widely used in quark pair creation model\cite{Micu:1968mk,Barnes:2005pb,Ferretti:2013faa}. The value of $(m_u/m_c)^2$ is about $1/20$  in the traditional quark model\cite{Barnes:2005pb,Ferretti:2013faa}. Hence, our assumptions of ${\cal B}(P_c\to J/\psi N)=10\%$ and ${\cal B}(P_c \to \pi N)=1\%$ are reasonable.

From Fig.~\ref{total}, it can be seen that the $\pi$ and $\rho$ meson exchanges provide a significant background contribution, while the two narrow bump structures come from $P_c^0(4380)$ and $P_c^0(4450)$ contributions. The thin bands for total cross sections indicate that the interference effects among the different contributions are extremely small. At c.m.~energy of $\rm{W}=4.38$ GeV and $4.45$~GeV, which regions we mainly concern, these effects are invisible and can be ignored. Hence, we only present the direct summations of each contribution in the following differential cross sections. With different $J^P$ assignments, the divergences among these total cross sections are small, which can hardly be used to identify the spin parities of the two $P_c$ states. The peaks, in the figure, are in the magnitude of 1 $\mu$b at the c.m.~energy $\rm{W}=4.38$ GeV and $4.45$~GeV, which can be measured in future high luminosity J-PARC experiments.

It is worthy mentioned that the contributions from neutral $P_c$ resonances are proportional to the branching ratios of $J/\psi N$ and $\pi N$ decay modes. In Ref.~\cite{Wang:2015jsa}, the low limit of $P_c \to J/\psi N$ ratio is assumed to be $5\%$. If the same low limit is employed, the present calculated total cross sections will reduce by a factor of 2, however the clear bump structures remain. For the cutoff parameter of $P_c$ states, a relatively small value is employed, which is more suitable for heavy meson production\cite{Wang:2015jsa,Kim:2011rm}. If this value increases, the contributions of two $P_c$ states will become larger. Actually, the form factor is approximate equal to 1 at resonance energy regions despite of the cutoff value, since $q^2-M_{Pc}^2\sim 0$. Our conclusions of the total and following differential cross sections remain while this cutoff parameter changes.

\begin{figure*}[htbp]
\includegraphics[scale=1.1]{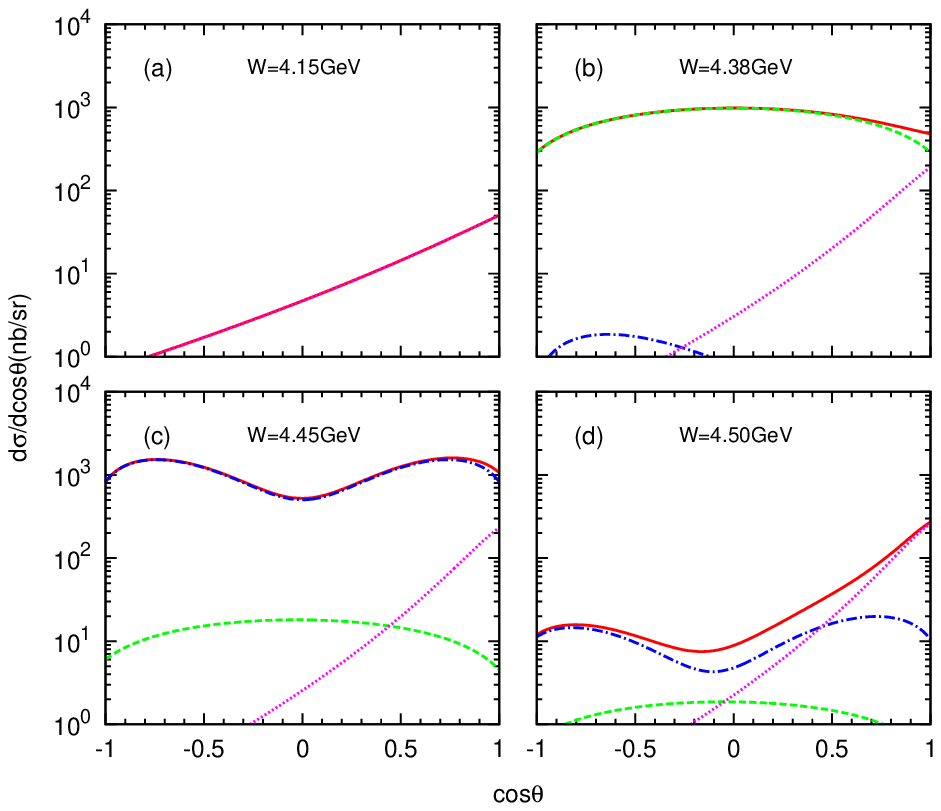}
\vspace{0.0cm} \caption{(Color online) The differential cross sections for $\pi^- p \to J/\psi n$ reaction at the c.m energies $\rm{W}=4.15$ GeV, 4.38 GeV, 4.45 GeV,  and 4.45 GeV. The $(P_c^0(4380),P_c^0(4450))$ corresponds to $(3/2^+,5/2^-)$ assumption. The red solid, green dashed, blue dot-dashed, and pink short dotted lines stand for total, $P_c^0(4380)$, $P_c^0(4450)$, and background contributions, respectively.}
\label{dif1}
\end{figure*}

\begin{figure*}[htbp]
\includegraphics[scale=1.1]{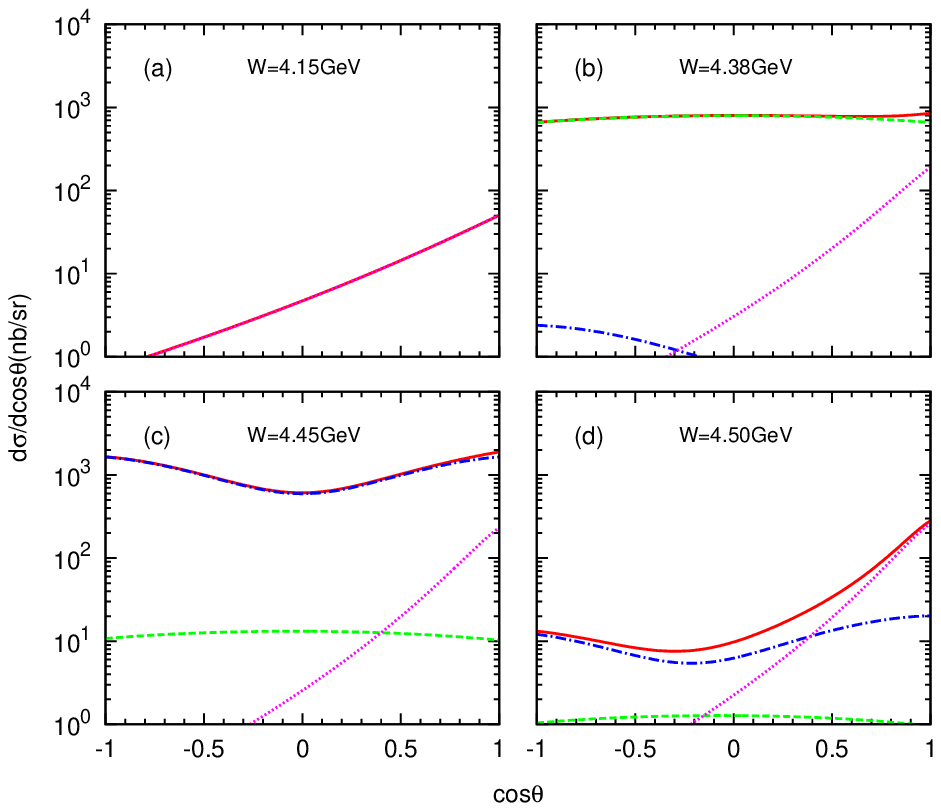}
\vspace{0.0cm} \caption{(Color online) The caption is the same as that of Fig.\ref{dif1}, but the $(P_c^0(4380),P_c^0(4450))$ corresponds to $(3/2^-,5/2^+)$ assumption.}
\label{dif2}
\end{figure*}

\begin{figure*}[htbp]
\includegraphics[scale=1.1]{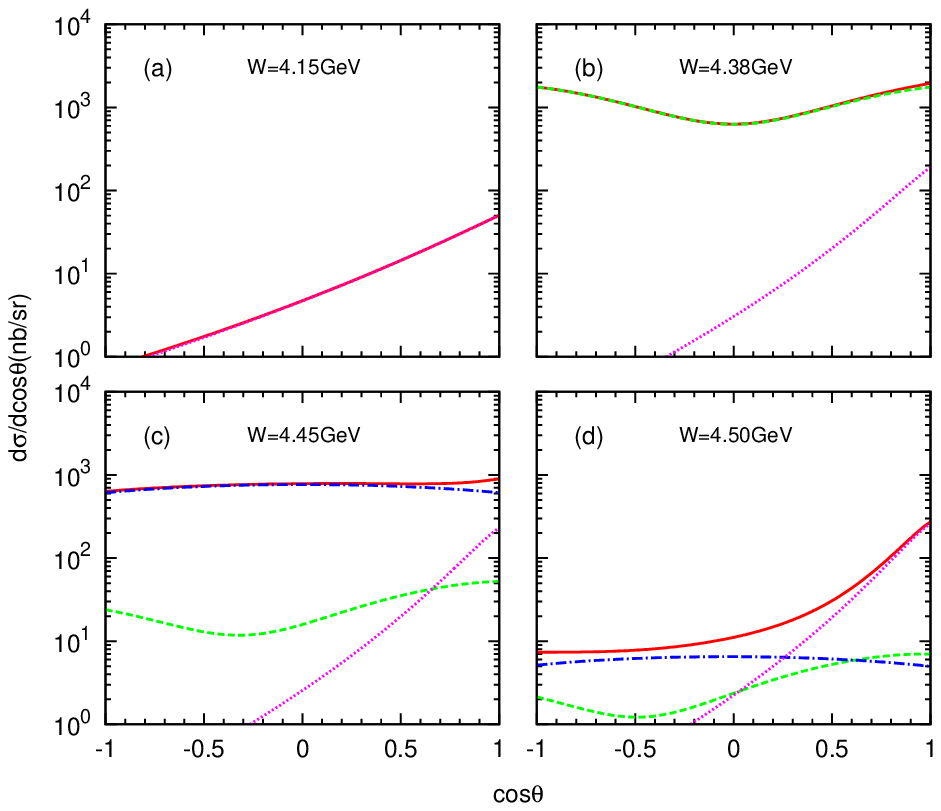}
\vspace{0.0cm} \caption{(Color online) The caption is the same as that of Fig.\ref{dif1}, but the $(P_c^0(4380),P_c^0(4450))$ corresponds to $(5/2^+,3/2^-)$ assumption.}
\label{dif3}
\end{figure*}

\begin{figure*}[htbp]
\includegraphics[scale=1.1]{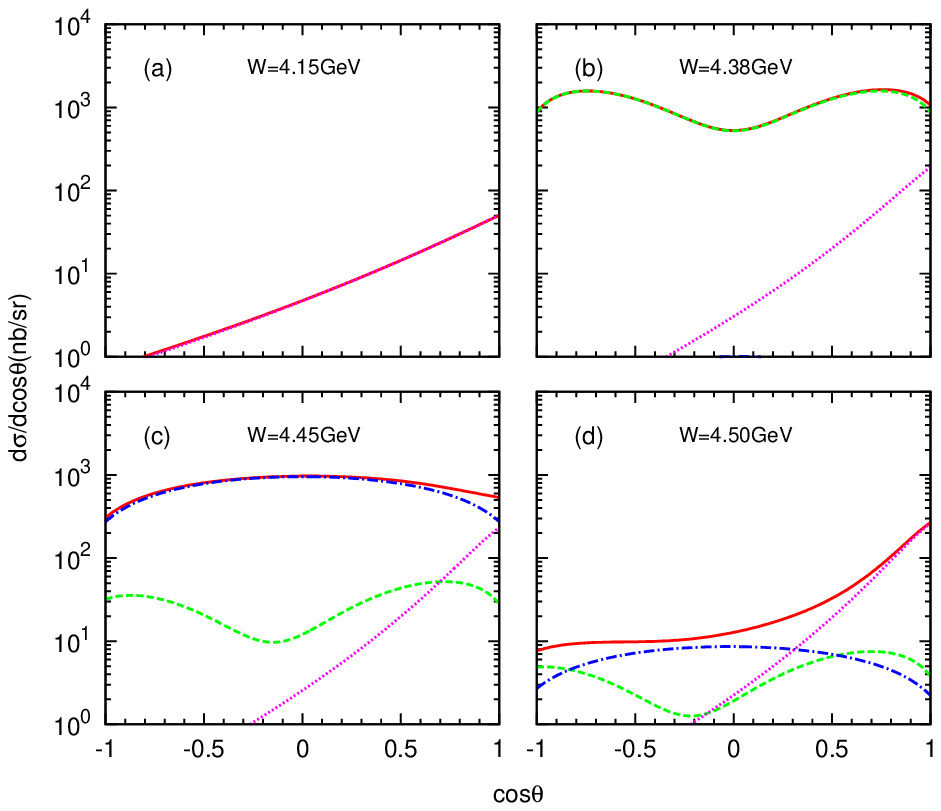}
\vspace{0.0cm} \caption{(Color online) The caption is the same as that of Fig.\ref{dif1}, but the $(P_c^0(4380),P_c^0(4450))$ corresponds to $(5/2^-,3/2^+)$ assumption.}
\label{dif4}
\end{figure*}

The differential cross sections at the c.m.~energies $\rm{W}=4.15$ GeV, 4.38 GeV, 4.45 GeV,  and 4.45 GeV are also presented in Fig.~\ref{dif1}-\ref{dif4}. It is shown that the $t$-channel meson exchanges provide forward contribution in the whole energy region and play a predominated role near the threshold. The differential cross sections at 4.38 GeV and 4.45 GeV are mainly from $P_c^0(4380)$ and $P_c^0(4450)$ contributions, respectively, which are also revealed by the total cross sections. The angular distributions of the two $P_c$ resonances are obviously different with forward background contribution and display significantly different behaviors with different $J^P$ assignments. It is expected that those specific features can be observed by future J-PARC experiments with high luminosity, and can help us to distinguish different spin parity assignments.

\section{Summary}{\label{summary}}

In this paper, the $\pi^- p \to J/\psi n$ reaction is studied, within an effective Lagrangian approach, in order to search for the neutral hidden charm pentaquark $P_c$ states. The background contribution mainly comes from $t$-channel $\pi$ and $\rho$ meson exchanges. For $s$-channel diagram, the $P_c$ states with different spin parity assignments are calculated and analyzed. We find that
the two states contribute clear bump structures in the total cross sections. Moreover, we also get that the differential cross sections of the $P_c$ states have significant divergences from background contribution and we explicitly show the different behaviors among the four spin parity assumptions. Those specific features of the angular distributions, together with the total cross sections with clear peak structures in the magnitude of 1 $\mu$b at c.m.~energy 4.38 GeV and 4.45 GeV, can be tested by future experiments in J-PARC.

\bigskip
\noindent
\begin{center}
{\bf ACKNOWLEDGEMENTS}\\
\end{center}

This project is supported by the National Natural Science Foundation of China under Grants No.~11475227, No.~11175220, No.~10975146, and No.~11475192. We acknowledge the Century Program of Chinese Academy of Sciences (Grant No.~Y101020BR0). The fund provided by the Sino-German CRC 110 ``Symmetries and the Emergence of Structure in QCD" project is also appreciated.

\end{document}